\documentclass[twocolumn,showpacs,floatfix,prl]{revtex4}

\usepackage[dvips]{epsfig}
\usepackage{amsmath}
\usepackage{amssymb}

\usepackage{float}

\begin{document}

\title{Majorana Fermions and Non-Abelian Statistics in Three Dimensions}

\author{Jeffrey C.Y. Teo and C.L. Kane}
\affiliation{Dept. of Physics and Astronomy, University of Pennsylvania,
Philadelphia, PA 19104}

\begin{abstract}
We show that three dimensional superconductors,
described within a Bogoliubov de Gennes framework can have zero
energy bound states associated with pointlike topological defects.
The Majorana fermions associated with these modes have non-Abelian
exchange statistics, despite the fact that the braid group is trivial
in three dimensions.  This can occur because the defects are
associated with an orientation that can undergo topologically
nontrivial rotations.  A feature of three dimensional systems is
that there are ``braidless" operations in which it is possible to
manipulate the groundstate associated with a set of defects without
moving or measuring them.  To illustrate these effects we analyze
specific architectures involving topological insulators and
superconductors.
 \end{abstract}

\maketitle

A fundamental feature of quantum theory is the quantum statistics
obeyed by identical particles.  For ordinary particles, Bose and
Fermi statistics are the only possibilities.  Emergent excitations
in correlated many particle systems, however, can exhibit fractional\cite{leinaas,wilczek,halperin,arovas}
and non-Abelian\cite{mr}
statistics.
The simplest non-Abelian excitations, known as Ising anyons\cite{review}, are Majorana fermion
states associated with zero energy modes that
occur in the Bogoliubov de Gennes (BdG) description of a paired condensate\cite{readgreen}.
They have been predicted in a variety of two dimensional (2D) electronic systems,
including the $\nu=5/2$ quantum Hall effect\cite{greiter}, chiral p-wave superconductors (SCs)
\cite{dassarma} and
SC-topological insulator (TI) structures\cite{fukane1}.  The groundstate of
$2N$ Ising anyons has a $2^N$ degeneracy, and when
identical particles are exchanged the state
undergoes a non-Abelian unitary transformation\cite{nayak,ivanov}.
Recent interest in non-Abelian statistics has been heightened by the
proposal to use these features for topological quantum computation\cite{kitaev1}.

Fractional and non-Abelian statistics are usually associated with
2D because in 3D performing an exchange twice is
topologically trivial.   In this paper, we show that in 3D Majorana
fermion states are associated with pointlike topological
defects, and that they obey non-Abelian exchange statistics, despite
the triviality of braids. Our motivation came from the study of 3D
SC-TI structures, where Majorana
fermions arise in a variety of ways, such as (i) vortices at
SC-TI interfaces\cite{fukane1}, (ii)
SC-magnet interfaces at the edge of a 2D TI\cite{kitaev1d,fukane2,nilsson}
and (iii) band inversion domain walls along a
SC vortex line.   While the Majorana fermions in these
cases can be identified using 1D or 2D effective theories, they
must occur in a more general 3D theory.  To unify them we
introduce a $\mathbb{Z}_2$ topological index that
locates the zero modes in a generic 3D BdG theory.  We then study
a minimal 8 band model in which the defects can be
understood as hedgehogs in a three component vector field.  Ising
non-Abelian exchange statistics arise because the hedgehogs have an {\it
orientation} that can undergo nontrivial rotations.  We will
illustrate the intrinsic three dimensionality of the Majorana states
by considering specific architectures involving SCs and
TIs.  A feature in 3D is the
existence of ``braidless" operations, in which the quantum
information encoded in the Majorana states can be manipulated without
moving or measuring\cite{bonderson} them.

To determine whether a Majorana mode is enclosed in a volume $V$
we topologically classify BdG Hamiltonians on $\partial V$,
the 2D surface $V$.  We
assume the Hamiltonian varies slowly, so we can consider adiabatic
changes as a function of two parameters ${\bf r}$ characterizing $\partial V$.
The problem is then to classify particle-hole (PH) symmetric BdG Hamiltonians ${\cal
H}({\bf k},{\bf r})$, where ${\bf k}$ is defined in a 3D
Brillioun zone (a torus $T^3$) and ${\bf r}$ is defined on a 2-sphere $S^2$.
PH symmetry is defined by an anti-unitary operator $\Xi$
satisfying $\Xi^2 = 1$ and ${\cal H}({\bf k},{\bf r}) = - \Xi {\cal H}(-{\bf
k},{\bf r}) \Xi^{-1}$.  Assuming no other symmetries, this corresponds to class D of the
general scheme\cite{schnyder,kitaev2}.   Since $V$ may or may not enclose a zero mode, we expect
a $\mathbb{Z}_2$ classification - a fact that can be established using
methods of K theory.

A formula for the topological invariant can be derived using a
method similar to Qi, Hughes and Zhang's\cite{qhz} formulation of the invariant
characterizing a 3D strong TI\cite{fukane0}.  We introduce a one parameter
deformation $\tilde {\cal H}(\lambda,{\bf k},{\bf r})$ that adiabatically connects
${\cal H}({\bf k},{\bf r})$ at $\lambda=0$ to a trivial Hamiltonian independent of
${\bf k}$ and ${\bf r}$ at $\lambda = 1$, while violating PH
symmetry.  PH symmetry can then be restored by including a mirror image
$\tilde {\cal H}(\lambda,{\bf k},{\bf r}) = - \Xi \tilde {\cal H}(-\lambda,-{\bf k},{\bf r}) \Xi^{-1}$
for $-1<\lambda<0$.  For $\lambda = \pm 1$, ${\bf k},{\bf r}$ can be replaced
by a single point, so the 6 parameter space $(\lambda,{\bf k},{\bf r})
\sim \Sigma(T^3 \times S^2)$ ($\Sigma$ denotes the suspension) has no boundary.
$\tilde {\cal H}$ defined on this space is characterized by its integer valued
third Chern character\cite{nakahara},
\begin{equation}
{\rm Ch}_3[{\cal F}] = {1\over {3!}} \left({i\over{2\pi}}\right)^3\int_{\Sigma(T^3\times S^2)}  {\rm
Tr}[ {\cal F}\wedge {\cal F}\wedge {\cal F} ].
\label{ch3}
\end{equation}
Here ${\cal F} = d{\cal A} + {\cal A} \wedge {\cal A}$ follows from
the non-Abelian Berry's connection ${\cal A}_{ij} = \langle u_i
|d|u_j\rangle$ associated with the negative energy eigenstates of $\tilde {\cal H}$.
Due to PH symmetry the contributions to (\ref{ch3}) for
$\lambda>0$ and $\lambda<0$ are equal.  Moreover, since it is a total derivative,
${\rm Tr}[{\cal F}^3] = d{\cal Q}_5$ (omitting the
$\wedge$'s), where
the Chern Simons 5-form is\cite{nakahara}
\begin{equation}
{\cal Q}_5 = {\rm Tr}\left[ {\cal A}\wedge (d{\cal A})^2 + (3/2){\cal A}^3 \wedge
d{\cal A} + (3/5) {\cal A}^5 \right].
\label{omega5}
\end{equation}
The integral over $\lambda>0$ can be then be pushed to the boundary $\lambda = 0$, so that
\begin{equation}
{\rm Ch}_3[{\cal F}] = {2\over {3!}}\left({i\over{2\pi}}\right)^3 \int_{T^3 \times S^2} {\cal Q}_5.
\label{boundary}
\end{equation}
A different deformation $\tilde{\cal H}$ can change ${\rm Ch}_3[{\cal
F}]$, but PH symmetry requires the change is
an {\it even} integer.  Likewise, the right hand side of (\ref{boundary}) is only gauge invariant up to an even
integer.  The parity of (\ref{boundary}) defines a $\mathbb{Z}_2$
topological invariant.  We write it as
\begin{equation}
\mu = \int_{S^2} \boldsymbol{\omega}({\bf r}) \cdot d{\bf A} \quad {\rm mod} \ 2
\label{mu}
\end{equation}
where the gauge dependent Chern-Simons flux $\boldsymbol{\omega} = (\omega^1,\omega^2,\omega^3)$ is
defined by integrating out ${\bf k}$,
\begin{equation}
{1\over 2}\epsilon_{ijk}\omega^i({\bf r}) dx^j\wedge dx^k =
{1\over 3} \left({i\over{2\pi}}\right)^3 \int_{T^3} {\cal Q}_5.
\label{omega}
\end{equation}
It is natural to associate $\mu$ with the presence of a zero mode
- a fact that will be checked explicitly below.

We now introduce a minimal model that leads to an appealing physical
interpretation for $\boldsymbol{\omega}({\bf r})$.  Since $\mu$ is
based on ${\rm Ch}_3[{\cal F}]$, we expect a minimum of 8 bands is
required. Consider a model parameterized by a three component vector
field ${\bf n}$ of the form, \begin{equation} {\cal H} =  -i\gamma_a
\partial_a + \Gamma_a n_a({\bf r}). \label{h} \end{equation} Here
$\gamma_a$ and $\Gamma_a$ ($a=1,2,3$) are $8 \times 8$ Dirac matrices
satisfying $\{\Gamma_a,\Gamma_b\}=\{\gamma_a,\gamma_b\}=2\delta_{ab}$
and $\{\Gamma_a,\gamma_b\}=0$.  $\cal H$ respects PH symmetry
provided $\Xi \Gamma_a \Xi^{-1}=-\Gamma_a$ and $\Xi \gamma_a \Xi^{-1}
= \gamma_a$. For ${\bf n}({\bf r}) = {\bf n}_0$, ${\cal H}$ has
eigenvalues $E({\bf k}) = \pm (|{\bf k}|^2 + |{\bf n}_0|^2)^{1/2}$,
so that for ${\bf n}_0 \ne 0$ there is a gap $2 |{\bf n}_0|$.   The
seventh Dirac matrix $\gamma_5 \equiv i \prod_a \gamma_a \Gamma_a$ is
not an allowed mass term because $\Xi \gamma_5 \Xi^{-1}=\gamma_5$.  A
more general Hamiltonian could also involve products of the Dirac
matrices, but such Hamiltonians can be homotopically deformed to the
form of (\ref{h}) without closing the gap\cite{kitaev2}.  To
regularize (\ref{h}) at $|{\bf k}|\rightarrow \infty$ we include an
additional term $\epsilon |{\bf k}|^2 \Gamma_3$, so ${\bf k}$ can be
defined on a compact Brillouin zone $S^3$.  The analysis is simplest
for $\epsilon \rightarrow 0$, where the low energy properties are
isotropic in ${\bf n}$.

${\cal H}$ can be physically motivated by considering a
BdG Hamiltonian describing ordinary and topological insulators coexisting with
superconductivity.  The Dirac matrices are specified by three sets of
Pauli matrices:  $\vec\tau$ for PH space,
$\vec\sigma$ for spin and $\vec\mu$ for an orbital degree of
freedom.  We identify $\vec\gamma = \tau_z \mu_z \vec \sigma$, $\Gamma_1 =
\tau_x$, $\Gamma_2 = \tau_y$ and $\Gamma_3 = \tau_z \mu_x$, along with
$\Xi = \sigma_y\tau_y K$.  ${\bf n}$ is then
$(\Delta_1,\Delta_2,m)$, where $\Delta=\Delta_1+i \Delta_2$ is a
SC order parameter and $m$ is a mass describing a band
inversion.  For $\Delta = 0$ (\ref{h}) is a doubled version of the
model for a 3D TI discussed in Ref. \onlinecite{qhz}.
For $\epsilon>0$, $m>0$ describes a trivial insulator, while $m<0$ describes a TI
with a band inversion near ${\bf k}=0$.  An interface where
$m$ changes sign corresponds to the surface of a TI, which
has gapless surface states.  Introducing $\Delta \ne 0$ to the interface then
describes the proximity induced SC state\cite{fukane1}.

To locate the zero modes, we take ${\bf n}({\bf r})$ to vary
slowly with ${\bf r}$ and evaluate $\boldsymbol{\omega}({\bf r})$ using (\ref{ch3}-\ref{omega}).
This can be done by noting that ${\cal H}$ defines a 6 component unit vector
given by the direction
$\hat{\bf d}^5({\bf k},{\bf r})$ of $(k_1,k_2,k_3,n_1+\epsilon |{\bf k}|^2,
n_2,n_3)$ on $S^5$.   The deformed Hamiltonian $\tilde {\cal H}$ can be defined by
adding $\lambda \gamma_5$ to ${\cal H}$, so that $\tilde{\cal H}$ defines
a vector on $S^6$ given by $\hat {\bf d}^6 = (\sqrt{1-\lambda^2} \hat
{\bf d}^5,\lambda)$.  ${\rm Ch}_3[{\cal F}]$ is the volume on $S^6$ swept out
by $\hat{\bf d}^6(\lambda,{\bf k},{\bf r})$.  Then $\int{\cal Q}_5$ is the volume in
the ``northern hemisphere" of $S^6$ swept out by $\hat{\bf d}^5({\bf k},{\bf r})$, which
is confined to the ``equator", $\lambda = 0$.  This is then related to
the area on $S^5$ swept out by $\hat{\bf d}^5({\bf k},{\bf r})$.  Performing the
integral on ${\bf k}$ for $\epsilon\rightarrow 0$ gives,
\begin{equation}
\omega^i({\bf r}) =  {1\over 8\pi}
\epsilon^{ijk} \hat{\bf n} \cdot \partial_j\hat{\bf n} \times \partial_k\hat{\bf n},
\label{omega3}
\end{equation}
where $\hat {\bf n} = {\bf n}/|{\bf n}|$.  Thus, the topological
charge that
signals a zero mode inside $V$
is the parity of the $S^2$ winding number of $\hat{\bf n}$ on $\partial V$.  Zero
modes are associated with {\it hedgehogs} in $\hat {\bf n}({\bf r})$.
A simple example of a hedgehog is a SC vortex at
the interface between a TI and an insulator.  Though the hedgehog
topological charge can be any integer, an even integer in (\ref{mu})
can be unwound by a ${\bf k}$ and ${\bf r}$ dependent gauge transformation.

The presence of a zero mode associated with a hedgehog can be
demonstrated with a simple linear model $n_a({\bf r}) = M_{ab}
r_b$, which has a hedgehog with charge ${\rm sgn}({\rm
det}[M])$ at ${\bf r}=0$.  This is solved by expressing $M$ in terms of
its principle axes: $M = {\cal O}_1^T {\rm diag}(M_1,M_2,M_3) {\cal O}_2$, where
${\cal O}_1$ and ${\cal O}_2$ are orthogonal matrices that diagonalize $MM^T$ and $M^TM$
respectively.  Defining $r'_a={\cal O}_{1ab}r_b$, $n'_a = {\cal O}_{2ab}n_b$,
$\gamma'_a = {\cal O}^T_{1ab}\gamma_b$ and $\Gamma'_a = {\cal O}^T_{2ab}\Gamma_b$, it is
straightforward to express ${\cal H}^2$ as three independent harmonic
oscillators,
\begin{equation}
{\cal H}^2 = \sum_a M_a( 2n_a + 1 - \xi_a)
\label{h2}
\end{equation}
where $n_a$ are oscillator quantum numbers and
$\xi_a = i \gamma'_a \Gamma'_a$ are commuting operators.  There is
a single zero energy state with $n_a = 0$ and $\xi_a = 1$.  This zero
mode is the non degenerate eigenstate with eigenvalue $3$
of $\sum_a \xi_a = i\sum_{a,b} \gamma_a {\cal O}_{ab}
\Gamma_b$, where ${\cal O} = {\cal O}_1 {\cal O}^T_2$.

A key feature of the zero mode is its dependence on
the relative {\it orientation} ${\cal O}$ of the principle axes in ${\bf r}$ and
${\bf n}$ space.  This can lead to a non trivial holonomy when $M$ (and hence ${\cal O}$)
varies.  We construct the zero mode by starting with
$|\Psi_0\rangle$ which satisfies $\sum_a i \gamma_a\Gamma_a
|\Psi_0\rangle = 3 |\Psi_0\rangle$ and then doing
a unitary transformation that takes $\Gamma_a$ to ${\cal O}_{ab}
\Gamma_b$.  If we parameterize the rotation ${\cal O}$ with a vector ${\bf\Omega}$
specifying the axis and angle $|{\bf\Omega}| \le \pi$,
then
\begin{equation}
|\Psi({\bf\Omega})\rangle = e^{\epsilon_{abc} \Gamma_a \Gamma_b \Omega_c/4} |\Psi_0\rangle.
\label{psio}
\end{equation}
This gauge satisfies
$\langle\Psi({\bf\Omega})|d \Psi({\bf\Omega})\rangle = 0$ for $|{\bf
\Omega}|<\pi$ but is not globally defined because
$|\Psi({\bf\Omega})\rangle = - |\Psi(-{\bf\Omega})\rangle$ when $|{\bf\Omega}| = \pi$.  This
reflects the nontrivial topology of $SO(3)$, characterized by the
homotopy $\pi_1(SO(3)) = \mathbb{Z}_2$.  When
${\cal O}$ varies along a nontrivial loop in $SO(3)$, the wavefunction of
the zero mode changes sign.  The associated Majorana operator $\gamma_i$
(not to be confused with the Dirac matrix $\gamma_a$) also changes sign.
This is one of our central results, and it is
this fact that allows Ising non-Abelian statistics in 3D.  A
simple example of a nontrivial loop is the $2\pi$ rotation that occurs when the
SC phase advances by $2\pi$.

Though we derived it with the linear 8 band model, our conclusion that there
is nontrivial holonomy for defects
is more general.  A formulation
based on (\ref{mu}) will appear elsewhere.  In the 8 band model,
a general defect history is characterized by ${\bf n}({\bf r},t)$, where ${\bf r}$ is
on a surrounding surface $S^2$ and $t$ varies on
a closed path $S^1$.  These are classified by the homotopy of
maps $S^2 \times S^1 \rightarrow S^2$, which were first analyzed by Pontrjagin\cite{pontrjagin},
and have appeared in other physical contexts\cite{jaykka}.
When the hedgehog's topological charge is $\pm p$ the classification is $\mathbb{Z}_{2p}$.
This is related to the integer Hopf invariant for maps
$S^3 \rightarrow S^2$ which applies when $p=0$.  Like the Hopf invariant, it
can be understood in terms of the linking of curves in $S^2 \times S^1$\cite{kapitanski}.
The discussion below resembles Wilczek and Zee's\cite{wilczek} analysis of the
statistics of skyrmions in the 2+1D nonlinear $\sigma$ model with
a Hopf term.

\begin{figure}
\centerline{ \epsfig{figure=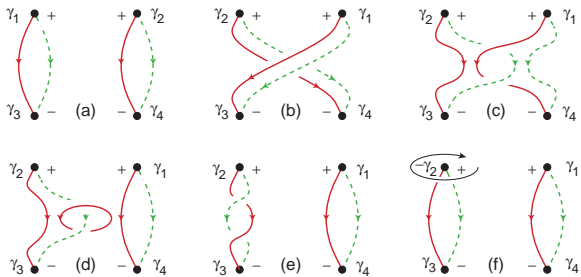,width=3.0in} }
 \caption{Inverse image paths depicting the interchange of two hedgehogs, as described in the text.
 (a-e) show a sequence of continuous deformations of ${\bf n}({\bf r})$ that preserve the orientation of the hedgehogs.
 In (f) a $2\pi$ rotation is required to return to the original configuration.}
\end{figure}

To study the exchange statistics of the Majorana modes
we consider the adiabatic evolution of the state when they are exchanged.
We thus consider
a 3+1D history $\hat {\bf n}({\bf r},t)$ satisfying
$\hat{\bf n}({\bf r},T)=\hat{\bf n}({\bf r},0)$
with hedgehogs at $\left\{{\bf r}_i(t)\right\}$ with ${\bf r}_{1(2)}(T)={\bf r}_{2(1)}(0)$.
To visualize $\hat {\bf n}({\bf r},t)$ it is useful to
consider the inverse image paths in ${\bf r}$
space that map to two specific points
on $S^2$.  Such paths begin and end on hedgehogs, and a crucial
role will be played by their linking properties.  Fig. 1
depicts four hedgehogs,
where the top two (positive) hedgehogs are interchanged.  At the
first step in (b) the locations of the hedgehogs have been
interchanged.  Since the inverse image paths have been ``dragged",
$\hat{\bf n}({\bf r})$ is not the same as it's original configuration.  Panels
(c-f) show a sequence of smooth deformations that untangle $\hat{\bf n}({\bf r})$.
The key point is that the two paths (which map to different
points on $S^2$) can never cross each other.  However, a path can
cross itself and ``reconnect", as in (c,d,e).  The deformations from
(a-e) preserve the orientation of the hedgehogs, but leave behind a
twist.  To return $\hat{\bf n}({\bf r})$ to its
original configuration in (f) requires a $2\pi$ rotation of one of
the hedgehogs.   This results in an interchange rule for the Majoranas,
\begin{equation}
T_{12}  : \quad \gamma_1 \rightarrow  \gamma_2 \quad ; \quad \gamma_2 \rightarrow
- \gamma_1
\label{tij}
\end{equation}
analogous the rules\cite{nayak,ivanov} for braiding vortices in 2D
and can be represented by
$T_{12}=\exp[\pi \gamma_1 \gamma_2/4]$.   Note that in (d,e) the twist could have been left on the
other side, which would have led to $T_{21}=T_{12}^\dagger$.
The two choices for $T_{12}$ correspond to
{\it physically distinct} interchange trajectories that generalize the
right and left handed braiding operations in 2D.

Performing the same interchange twice leads to a nontrivial operation,
since $T_{ij}^2 = \gamma_i\gamma_j$ changes the sign of
both $\gamma_i$ and $\gamma_j$.  This is natural in 2D
because it is a non contractable braid.  In 3D, however,
$T_{ij}^2$ can
be smoothly deformed into an operation in which all particles are held
fixed.
Thus there is an operation, specified by a history
$\hat{\bf n}({\bf r},t)$, that rotates any pair of {\it stationary} hedgehogs by
$2\pi$, and implements the operation $\gamma_i\gamma_j$.
The existence of such ``braidless" operations is a feature of Ising non-Abelian statistics in
3D.  Although these operations form an admittedly
limited Abelian subgroup, they nonetheless offer a method for manipulating the quantum information
encoded in the Majorana fermions without moving or measuring them.

We now illustrate these effects using specific
architectures involving TIs and SCs.  It is
easiest to engineer Majorana modes using structures involving interfaces or vortex
lines, where $\boldsymbol{\omega}$ is confined to lines or planes.
Nonetheless, such structures can exhibit intrinsically 3D effects.
Consider first the structure in Fig. 2(a), which involves two
disconnected spherical TIs surrounded by a
SC and connected to each other by a Josephson junction.
Suppose that each sphere has a single $\pm$ pair of vortices, so that
there are 4 Majorana states on the spheres.
The internal state of the Majorana fermions can be represented in
a basis of eigenstates of $n =
i\gamma_1\gamma_2$ and $n' = i \gamma_3\gamma_4$.  For an isolated system
the parity of $n+n'$ is fixed, so the system is a
single qubit with basis vectors $|nn'=00,11\rangle$.  The state can be initialized
and measured in this basis with a probe that couples to both $\gamma_1$ and
$\gamma_2$.  Suppose $\gamma_3$ is adiabatically
transported around $\gamma_1$ as shown.  This is similar to a 2D braid, and
it implements
$T^2_{13}$, which interchanges $|00\rangle$ and $|11\rangle$.  Note,
however, that this ``braid" can be smoothly contracted to zero by
sliding the path around the other side of the sphere.  In the process,
however, the path crosses the junction connecting the
spheres, resulting in a $2\pi$ phase slip.  Thus the braiding operation can be smoothly deformed into a
``braidless" operation, where $\gamma_1$,$\gamma_3$ are held fixed, but the
phase difference between the two SCs advances by $2\pi$.

\begin{figure}
\centerline{ \epsfig{figure=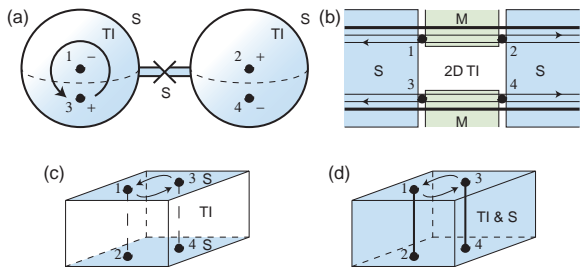,width=3.0in} }
 \caption{Architectures demonstrating 3D Majorana states using SCs and TIs.
 (a) Braiding 3 around 1 can be deformed into a braidless operation.  (b) A geometry for implementing
 braidless operations with a Josephson junction device\cite{fukane2}.   (c,d) Thin film geometries for interchanging
and measuring Majorana states. (c) shows a layered SC-TI-SC structure, while (d) shows a thin film of a
SC weakly doped TI.}
\end{figure}

A similar but more feasible version of the braidless operation
occurs for the Josephson junction structure in Fig. 2b,
which involves Majorana modes at a SC-magnet interface at the edge of a
2D TI.
As argued in Ref. \onlinecite{fukane2}, when the phase difference
across the junction is advanced by $2\pi$, the fermion parity
associated with $i\gamma_1\gamma_2$ changes, resulting in a fractional Josephson
effect.

Fig. 2c shows a TI coated on the top and bottom with
SC films.  In a magnetic field, Majorana states occur
at vortices on both the top and bottom.  If the SC films
are thinner than the penetration depth the field is constant, so the top and bottom vortices are
independent.  If the TI is
thin, there will be a weak vertical coupling that splits nearby Majorana
modes according to $n = i\gamma_1\gamma_2$ and $n' = i\gamma_3\gamma_4$.  The states
$|nn'\rangle$ will have slightly different charges, which
may allow $n$, $n'$ to be measured with a sensitive
charge detector.  Suppose the state is initially $|00\rangle$.
Interchanging $\gamma_1$, $\gamma_3$ on the
top keeping the bottom fixed leads to the entangled state
$(|00\rangle + |11\rangle)/\sqrt{2}$.
A variant on this geometry (Fig. 2d) is a thin film of a
bulk SC {\it weakly doped} TI, in which
the surface states acquire SC similar to
the proximity induced state.  In this
case, the interchange of $\gamma_1$, $\gamma_3$ involves a reconnection of the vortex
lines (which in principle have a finite energy gap) connecting them.

Recently, SC has been
observed in Cu$_x$Bi$_2$Se$_3$ for $x \sim .15$\cite{ong}.  It will be interesting to determine
whether this material is in the weakly doped regime, with Majorana
modes at the {\it ends} of vortex lines, or a more conventional
SC, which could be used in Fig. 2c.
There are certainly technical challenges associated
with manipulating the vortices and measuring their charge state.
Nonetheless, we hope that
the prospect of detecting 3D non-Abelian statistics in
such a system will provide motivation for further exploration.

We thank Liang Fu for insightful discussions and Bryan Chen and Randy Kamien for
introducing us to the Pontrjagin invariant.  This work was supported by NSF grant 0906175.

\end{document}